\documentclass[12pt]{article}
\usepackage[cp866]{inputenc}
\usepackage {amssymb}
\usepackage {amsmath}
\usepackage {graphicx}
\usepackage{setspace}
\input epsf

\input amssym.def
\input amssym.tex

\begin{document}

\begin{spacing} {1.5}

\begin{center}
{\bf Isotropic matrix elements of the Boltzmann equation collision integral.}
\end{center}

\begin{center}
I.A.Ender**, L.A.Bakaleinikov*, E.Yu.Flegontova*, A.B.Gerasimenko*
\end{center}

{\it* Ioffe Institute, Russia}

{\it** Saint-Petersburg State University, Russia}

{\footnotesize e-mail: fl.xiees@mail.ioffe.ru}

\vspace{12pt}

%Предложен алгоритм построения матричных элементов интеграла столкновений изотропного %по скоростям нелинейного уравнения Больцмана Эти матричные элементы используются в %качестве стартовых в рекуррентной процедуре расчета матричных элементов %неизотропного по скоростям интеграла столкновений, изложенной в нашей предыдущей %работе.
We propose an algorithm for calculating  matrix elements of the non-linear Boltzmann equation collision integral in isotropic case. These matrix elements are used as starting  ones in the  recurrence procedure for calculating the matrix elements of the collision integral, which is non-isotropic with respect to velocities, as described in our previous paper.
% Кроме того, изотропные матричные элементы представляют и самостоятельный интерес %для расчета изотропной релаксации в ряде задач физико-химической кинетики. 
In addition, isotropic matrix elements are of independent interest for the calculation of isotropic relaxation in a number of problems of  physical-chemical  kinetics.
% Показано, что коэффициенты разложения изотропных матричных элементов по $\Omega$-%интегралам связаны рекуррентными соотношениями, позволяющими построить процедуру их % последовательного определения. 
 It is shown that the coefficients of the  isotropic matrix elements  expansion in terms of  $ \Omega $ -integrals are connected  by recurrence relations that allow us to  calculate them sequentially.

\section{Introduction}

%Развитие современной техники  за счет миниатюризации устройств и за счет перехода к %более напряженным режимам работы требует решения кинетических задач на уровне %функции распределения (ФР). 
The progress  of modern technology both due to miniaturization of devices and  more intense operating modes requires  to treat kinetic problems  by consideration  of the distribution function (DF).
%Информация о ФР в области больших скоростей оказывается определяющей и для расчета %физико-химических процессов в газе и плазме.
Knowledge of the distribution function at high values of the velocities turns out to be decisive for the calculation of   physical-chemical  processes in gas and plasma. %Наконец, к необходимости разработки методов расчета ФР в сильно неравновесной %ситуации приводит и естественное развитие самой кинетической теории.
Finally, the natural evolution of the kinetic theory itself leads to the necessity of developing methods for calculating the distribution function  that is far from equilibrium.
% Одним из методов решения уравнения Больцмана при сильном отклонении от равновесия %является 
%метод полиномиальных разложений, в частности, моментный метод, в котором ФР %раскладывается по функциям Барнетта
%$H_{r,l,m}=S^r_{l+1/2}(x) Y_{l,m} (\theta, \phi)$, где $S^r_{l+1/2}(x)$ -- полиномы %Сонина, 
%$Y_{l,m} (\theta, \phi)$ -- сферические гармоники. 
One of the methods for solving the Boltzmann equation for a  system that deviates strongly from equilibrium is the method of polynomial expansions, in particular, the moment method in which the distribution function is expanded in terms of Barnett functions $H_{r,l,m}=S^r_{l+1/2}(x) Y_{l,m} (\theta, \phi)$, where $S^r_{l+1/2}(x)$ are Sonine polinomials.

% Основной причиной, сдерживающей развитие этого метода, было отсутствие  %достаточного количества матричных элементов интеграла столкновений (МЭ). 
 The main  drawback of this method was the lack of  sufficient number of computed collision integral matrix elements (ME).
%В кинетической теории газов, как правило, вычислялись  неизотропные  МЭ с  малыми %индексами $ l $ \cite{ChK,FK}. 
As a rule, non-isotropic ME  with small indices $l$ were calculated in the kinetic theory of gases \cite{ChK,FK}.
%Это связано c их исключительной ролью при вычислении коэффициентов переноса при  %малых отклонениях от равновесия.
This is due to their  great importance for the calculation of the transport coefficients  in the case of small deviations from equilibrium.
%В сильно неравновесных процессах для аккуратного построения ФР необходимо знание МЭ %с большими индексами. 
In highly nonequilibrium processes, accurate calculation of the distribution function requires matrix elements with large indices.
%Существующие прямые аналитические формулы для определения МЭ (см. например, %\cite{Viehland78})  чрезвычайно громоздки. 
The  analytical formula for ME (see, for example, \cite{Viehland78}) are extremely cumbersome.
% В  \cite{book, ender2007} был  предложен значительно более простой способ %последовательного определения МЭ с помощью рекуррентных процедур.
In \cite{book, ender2007}, a much simpler method  to determine ME sequentially using recurrence procedures has been proposed.

%Настоящая  работа тесно связана с нашей предыдущей работой \cite{RecME1}, в которой %описана такая рекурретная процедура построения  любых неизотропных  (соответствующих %неизотропной по скоростям ФР) матричных элементов.
The present paper is closely related to our previous work \cite{RecME1}, which describes the  recurrence procedure for constructing any non-isotropic (corresponding to a velocity-non-isotropic distribution function) matrix elements.  
%При этом в качестве стартовых предлагается использовать изотропные МЭ.
 The isotropic ME are used in the procedure as starting ones. 
%Важно отметить, что сведения о сечениях взаимодействия используются только на этапе %построения изотропных МЭ.
It is important to note that information about interaction cross sections is  used only at the stage of isotropic ME calculation.
%В статье будет показано, что любые  изотропные МЭ как линейные, так и нелинейные  %можно представить  в виде линейной комбинации $\Omega$-интегралов, коэффициенты %которых связаны алгебраическими рекуррентными соотношениями.
It will be shown in the paper that any isotropic MEs, both linear and non-linear, can be represented as a linear combination of $ \Omega $ -integrals whose coefficients are related by algebraic recurrence relations.

%Следует отметить, что построение изотропных МЭ важно не только для завершения нашей %рекуррентной процедуры.
It should be noted that the calculation of isotropic ME is important not only to complete our recurrent procedure.
%Сравнительно недавно появилась статья Шизгала и Дриди  \cite{SD}, в которой %вычисляются коэффициенты разложения  линейных изотропных МЭ по $ \Omega $ - %интегралам.
 Recently, an article by Shizgal and Drydy \cite {SD} appeared, in which the expansion coefficients of linear isotropic MEs in terms of $ \Omega $ integrals are calculated.
% В этой статье описывается целая серия прикладных задач, для которых важно знание %линейных изотропеых МЭ.
 This article describes  a number of applied problems  where linear isotropic ME  are necessary.
%  Это важно, например, при расчете ФР небольшой примеси атомов  (или ионов) на фоне %равновесно  распределенных атомов другого сорта.
  This is important, for example, in the calculation of a  distribution function of small quantity of the atoms (or ions)  in the equilibrium gas of atoms of another kind.     

%В \cite{SD} подчеркивается роль изотропной релаксации, т.е. этапа, на котором ФР %релаксирует, оставаясь изотопной. Скорости физико-химических превращений на этом %этапе определяются поведением ФР в области больших скоростей. 
In \cite{SD}, the  importance of the isotropic relaxation, i.e. stage at which the DF relaxes, remaining isotopic  is emphasized. The rates of  physical-chemical processes physical-chemical transformations at this stage are determined by the behaviour of the DF  at high velocities.

%Итак, разложение  изотропных матричных элементов по $ \Omega $-интегралам важно %как с точки зрения получения стартовых МЭ  для применения  рекуррентных  процедур, %так и само по себе, для решения ряда релаксационных задач, необходимых, в частности, %для физико-химической кинетики.
Thus, the expansion of isotropic matrix elements in $ \Omega $-integrals is important both  to obtain starting ME for  the recurrence procedure and, as itself,  to solve  relaxation problems, in particular, for  physical-chemical kinetics.

\section{Statement of the problem}

%Нелинейные интегралы столкновений  для каждой пары компонентов газовой смеси  с %массами $m_a$ и $m_b$ и плотностями $n_a$ и $ n_b $ в случае 
%изотропных по скоростям ФР ($f_a({\bf v})=f_a(v)$) имеют вид  
Non-linear collision integrals for each pair of gas mixture components with masses $ m_a $ and $ m_b $ and densities $ n_a $ and $ n_b $  for isotropic distribution function ($ f_a ({\bf v}) = f_a (v) $ ) have the form:
\begin{equation}
\hat{I}(f_a,f_b)=n_an_b\int \left(f_a({ v}_1)f_b({ v}_2) - f_a({ v})
f_b({ v}') \right) g\sigma_{ab} (g,\theta)\ d{\bf v}'\ d{\bf k}.
\label{eq1new}
\end{equation}
%
%Cкорости частиц до и после столкновения связаны соотношениями
The particle velocities before and after the collision are related by
\begin{displaymath}
{\bf v}_1={\bf v}_0 -\mu_b{\bf  k}g,\qquad    {\bf  v}_2={\bf  v}_0+\mu_a{\bf
k}g,\qquad   {\bf v}_0=(\mu_a {\bf v}+{\mu_b \bf v}'),
\end{displaymath}
\begin{equation}
\mu_a=m_a/(m_a+m_b), \qquad   \mu_b=m_b/(m_a+m_b) ,   
\label{eqmass}
\end{equation}
\begin{equation}
{\bf g} = {\bf v}_1 - {\bf v}_2, \quad {\bf  g}'={\bf  v}-{\bf  v}', \quad {g} = {g}' ,
\nonumber
\end{equation}
%
%где ${\bf k}$ ---  единичный  вектор,  направленный вдоль ${\bf g}$. 
where ${\bf k}$ is unit vector, directed along ${\bf g}$. 
%Угол рассеяния $\theta$ определяется соотношением $\cos\theta={\bf k}  \cdot  {\bf   %g'}/g$, а $\sigma_{ab} (g,\theta)$  ---  дифференциальное сечение рассеяния.
The scattering angle $ \theta $ is determined by the relation $ \cos \theta = {\bf k} \cdot {\bf g '} / g $, and $ \sigma_ {ab} (g, \theta) $ is the differential scattering cross section.
%

%При применении моментного метода к изотропным по скоростям задачам ФР  $a$-той %компоненты смеси $ f_a(v) $ разлагают по полиномам Сонина $S_{1/2}^r(x)$  с %максвелловским весом $ M_a$ 
When applying the moment method to isotropic problems, the distribution function of the $a$-th component of the mixture $ f_a (v) $ is expanded in Sonin polynomials $ S_ {1/2}^r (x) $ with the Maxwellian weight $ M_a $
\begin{equation}
f_a(v,t) = M_a \sum^{\infty}_{r=0} C^a_{r}(t)S^{r}_{1/2}\left( \frac{m_a v^2}{2kT}\right),\qquad
M_a = \left({m_a\over 2kT\pi}\right)^{3/2}\exp\left({-m_a v^2
\over 2kT}\right) .
\label{EOP11}
\end{equation}
%Здесь  $k$ --- постоянная Больцмана, $T$ --- температура весового максвеллиана.
Here $ k $ is the Boltzmann constant, and $ T $ is the temperature of the weighted Maxwellian.
The isotropic matrix elements $ (K^a_{a,b})^{r}_{r_1,r_2} $ are defined in terms of the collision integrals $ {\hat I} (f_a, f_b) $ as follows:
\begin{equation}
(K^a_{a,b})^{r}_{r_1,r_2} = \frac{4\pi}{\nu_{r}} \left( \int^{\infty}_{0} S^{r}_{1/2}(c_a^2)
\hat{I}_{a,b}(M_a S^{r_1}_{1/2}(c_a^2), M_b S^{r_2}_{1/2}(c_b^2))v^2\ dv\right),
\label{EOP1.3}
\end{equation}
\begin{equation}
{c}_a=\left({m_a\over 2kT}\right)^{1/2}{ v}, \ 
{ c}_b=\left({m_b\over 2kT}\right)^{1/2}{ v},\quad
\nu_{r} = {(2r + 1)!!\over (2r)!!}.
\label{EOP1.4}
\end{equation}
%

%В более общем, неизотропном случае МЭ определяются формулой, %аналогичной(\ref{EOP1.3}), однако в качестве базисных функций вместо полиномов %Сонина используются функции Барнетта
In a more general, non-isotropic case, ME are defined by a formula similar to (\ref{EOP1.3}), however,  Barnett's functions  $H_{r,l,m}=S^r_{l+1/2}(x) Y_{l,m} (\theta, \phi)$ are used as the basis functions instead of Sonine polynomials,
%, где $Y_{l,m} (\theta, \phi)$ -- сферические гармоники.  
% где $Y_{l,m} (\theta, \phi)$ -- сферические гармоники.
where $Y_{l,m} (\theta, \phi)$ are spherical harmonics.
 
%Как показано в \cite{PhFl,book}, изотропные МЭ  $(K^a_{a,b})$ с различными  %индексами связаны соотношением
As it was shown in \cite{PhFl, book}, isotropic ME $(K ^ a_ {a, b})$ with different indices are related by
\begin{displaymath}
T {d(K^a_{a,b})^{r}_{r_1,r_2}(T)\over dT} - (r_1+r_2-r)(K^a_{a,b})^{r}_{r_1,
r_2}(T) =
\end{displaymath}
\begin{equation}
=r(K^a_{a,b})^{r-1}_{r_1,r_2}(T) - (r_1+1)(K^a_{a,b})^{r}_{r_1+1,r_2}(T) -
 (r_2+1)(K^a_{a,b})^{r}_{r_1,r_2+1}(T).
\label{E0P1.37}
\end{equation}
%%%  Number: 1.36  %%%

%В случае степенной зависимоти потенциала $\varphi$ от расстояния $r$ ($ \varphi \sim %1/r^{\kappa} $) сечение рассеяния представляется в виде 
In the case of a power-law dependence of the potential $\varphi$ on the  distance $r$ ($ \varphi \sim 1/r^{\kappa} $) the cross section is represented in the form
\begin{equation}
\sigma_{ab}(g,\theta)=g^{\gamma-1}F^{ab}_\gamma(z) .
%=\left({2kT\over m_{ab}}\right)^
%{(\gamma-1)/2}\bar{g}^{\gamma-1}F^{ab}_\gamma(\theta).
\label{ppsigma} 
\end{equation}
%Здесь $z=sin^2(\theta/2)$, $\gamma = (\kappa - 4)/{\kappa}$, $ F^{ab}_\gamma(\theta) %$ -- угловая часть сечения, $\theta$ -- угол рассеяния. 
Here $z=sin^2(\theta/2)$, $\gamma = (\kappa - 4)/{\kappa}$, $ F^{ab}_\gamma(\theta) $ is the angular part of cross section, $\theta$ is the scattering angle.

%В случае сечения рассеяния (\ref{ppsigma}) МЭ пропорционален $T^{\mu}$, где 
% $\mu = \gamma / 2$ (см., например, \cite{JTP1994}). 
In the case of scattering cross section (\ref{ppsigma}) the ME is proportional to $T^{\mu}$, where $\mu = \gamma / 2$ (see, e.g. \cite{JTP1994}). 

%Cоотношения (\ref{E0P1.37}) становятся алгебраическими и  представляют собой %рекуррентные соотношения: 
Relations (\ref{E0P1.37}) become algebraic recurrence relations: 
\begin{displaymath}
(\mu - r_1 - r_2 + r)(K^a_{a,b})^{r}_{r_1,r_2}(T) =
\end{displaymath}
\begin{equation}
=r(K^a_{a,b})^{r-1}_{r_1,r_2}(T) - (r_1+1)(K^a_{a,b})^{r}_{r_1+1,r_2}(T) -
 (r_2+1)(K^a_{a,b})^{r}_{r_1,r_2+1}(T).
\label{recrelp}
\end{equation}
% 
%Существование рекуррентных соотношений позволяет построить рекуррентную процедуру, в %которой стартовыми являются линейные МЭ $ K^{r}_{r_1,0} $ (первый тип) или $ K^{r}%_{0,r_2} $ (второй тип).
The relations (\ref{recrelp}) allow us to establish the recurrence procedure for sequential determination of ME and 
linear ME $ K^{r}_{r_1,0} $ (first type) or $ K^{r}_{0, r_2} $ (second type) are the starting ones.
% В случае простого газа для этих стартовых МЭ получены простые аналитические формулы %\cite{PhFl,book}. 
The simple analytical formulas for these starting ME were obtained \cite{PhFl, book} in the case of a monoatomic gas. 
%На основе рекуррентных соотношений (\ref{recrelp}) нами была реализована программа %расчета МЭ с большими индексами (с индексами порядка сотни и более), для чего %оказалось необходимым разработать специальные процедуры точной арифметики. (Нарушена логика с высокой точностью!!!)
On the basis of recurrence relations (\ref {recrelp}) we have created a program for calculating ME with large indices (with indices of the order of hundreds or more), for which it turned out necessary to develop special procedures  to perform calculations with large amount of significant digits.
%Это позволило в изотропном случае решать нелинейное уравнение Больцмана 
%и рассчитывать функцию распределения вплоть до 8--10 тепловых скоростей.
This made it possible to solve the non-linear Boltzmann equation in the isotropic case and calculate the distribution function in the range of up to 8-10 thermal velocities.
% Для максвелловских молекул расчет проводился до 12 тепловых скоростей.
 For Maxwellian molecules, calculations were made up to 12 thermal velocities.
%  Было проведено сравнение с аналитическим решением БКВ \cite{BKW} и получено полное %совпадение. При этом рассматривалось отношение ФР к равновесному максвеллиану.
  The comparison with the analytical solution of the BKW \cite{BKW} was made and the complete coincidence was observed. In this case, the ratio of the distribution function to the equilibrium Maxwellian was considered.  
 
%В случае произвольных законов  взаимодействия в \cite{ChK,FK} аналитически проведен %расчет линейных неизотропных МЭ при малых значениях индексa  $l$ ($l=1,2 $), %отвечающего за разложения ФР по сферическим гармоникам. 
In the case of arbitrary interaction laws, the calculation of linear nonisotropic MEs for small values of the index $ l $ ($ l = 1,2 $), corresponding to the expansion of the DF in spherical harmonics, is analytically carried out in \cite{ChK, FK}.
%В результате такие  МЭ   представляются  в виде  линейной комбинации двойных %интегралов ($\Omega$-интегралов),  содержащих сечение взаимодействия
% $\sigma_{ab} (g,\theta)$. 
As a result, such MEs are represented as a linear combination of double integrals ($ \Omega $ -integrals) containing the cross section of the interaction $ \sigma_{ab} (g, \theta) $.

%Следуя \cite{FK}, запишем $\Omega$-интегралы в виде 
Following \cite {FK}, we write $ \Omega $ -integrals in the form
\begin{equation}
\Omega ^{(n,m)}_{ab}=\left({kT\over 2\pi m_{ab}}\right)^{1/2}\int
^\infty_0 e^{-\bar{g}^2}\bar{g}^{2m+3} Q^{(n)}_{ab} \ d\bar{g},
\label{EOP2.6} 
\end{equation}
%где $ m_{ab}=(m_am_b)/(m_a+m_b)$ -- приведенная масса пары молекул сортов $a$ и $b$, %а безразмерная переменная интегрирования $\bar {g}=g \sqrt{m_{ab}/2kT} $. 
where $ m_ {ab} = (m_am_b) / (m_a + m_b) $ is the reduced mass of a pair of molecules  $ a $ and $ b $, and the dimensionless integration variable $ \bar {g} = g \sqrt {m_ {ab } / 2kT} $.
%Величина $Q_{ab}^{(n)}$ в (\ref{EOP2.6})определяется выражением 
The value of $ Q_ {ab} ^ {(n)} $ in (\ref{EOP2.6}) is determined by the expression
\begin{equation}
Q^{(n)}_{ab}=2\pi \int ^\pi _0\big(1-\cos ^{n}\theta \big)
\sigma_{ab} (g,\theta ) \sin \theta \ d\theta.
\label{E0P2.10} 
\end{equation}
%
%В \cite{preprint} по аналогии с хорошо известным из кинетической теории методом %расчета интегральных скобок \cite{ChK,FK} получены аналитические  представления %изотропных линейных МЭ как первого, так и второго типов в виде конечных сумм  
% $\Omega$-интегралов. 
Analytic representations of isotropic linear MEs of both first and second types in the form of finite sums of $ \Omega $ -integrals are obtained in \cite{preprint} by analogy with the method of calculating integral brackets \cite{ChK, FK}, which is well known from the kinetic theory.
%Отметим, что в \cite{ChK,FK} рассматривались 
%только те линейные неизотропные МЭ, которые необходимы для вычисления 
%коэффициентов переноса. 
%При этом, так же как и в \cite{ChK,FK}, использована производящая функция для %полиномов Сонина, выбираются те же переменные интегрирования, так же проводится %изменение порядков многочисленных суммирований. В изотропном случае расчеты   %оказались значительно более простыми.
In this case, just as in \cite{ChK, FK}, the generating function for Sonine polynomials is used, the same integration variables are chosen, and the orders of numerous summations are changed  in the same way. In the isotropic case, the calculations turned out to be much simpler.
% Особенно просто через  $\Omega$-интегралы выражаются  линейные изотропные МЭ %второго типа $ K^{r}_{0,r_2} $, поэтому имеет смысл выбирать коэффициенты разложения %именно этих МЭ в качестве стартовых в рекуррентной процедуре получения всех %изотропных МЭ. 
 It is especially simple to express linear isotropic ME of the second type $ K^{r}_{0, r_2} $ through $ \Omega $ -integrals.  Therefore we choose the expansion coefficients of these ME as the starting ones to calculate all isotropic ME by recurrence procedure.
%При этом удобно вместо функций $Q_{ab}^{(n)}$ использовать
In this case, instead of the functions $ Q_{ab}^{(n)} $, it is convenient to use
\begin{equation}
Q^{(n)}_{*ab}=2\pi(1-\delta_{n,0})\int^{\pi}_{0} \big(1-\cos \theta \big)^n
\sigma_{ab} (g,\theta )\sin \theta \ d\theta .
\label{E0P2.16} %! - ML - !%
\end{equation}
%
%По аналогии с  $\Omega$-интегралами определим $\Omega_*$-интегралы
By analogy with $ \Omega $ -integrals, we define $ \Omega_* $ - integrals
\begin{equation}
\Omega ^{(n,m)}_{*ab}=\left({kT\over 2\pi m_{ab}}\right)^{1/2}\int
^\infty_0 e^{-\bar{g}^2}\bar{g}^{2m+3} Q^{(n)}_{*ab} \ d\bar{g}.
\label{E0P2.17} %! - ML - !%
\end{equation}
%
% $\Omega_*$-интегралы легко выражаются через $\Omega$-интегралы, и
%наоборот:
$ \Omega_* $ - integrals can be easily expressed in terms of $ \Omega $ -integrals, and conversely:
\begin{equation}
\Omega ^{(n,m)}_{*ab}=\sum^{n}_{l=0}(^{n}_{l})(-1)^{l+1}
\Omega ^{(l,m)}_{ab},
\quad
\Omega ^{(n,m)}_{ab}=\sum^{n}_{l=1}(^{n}_{l})(-1)^{l+1}
\Omega ^{(l,m)}_{*ab}.
\label{E0P2.21a} 
\end{equation}
%
%Приведем здесь полученные в \cite{preprint} представления линейных изотропных МЭ %первого и второго типов через $\Omega_*$-интегралы. Для МЭ первого типа имеем
We give here the representations of linear isotropic ME of the first and second types obtained in \cite{preprint} via $ \Omega_* $ - integrals. For ME of the first type, we have 
%В \cite{preprint} получено представление МЭ $(K^a_{a,b})^q_{0,p}$ через $\Omega_*$-%интегралы 
%в виде

%Для сравнения приведем полученное в \cite{preprint} выражение МЭ первого типа 
% $(K^a_{a,b})^q_{p,0}$ через $\Omega_*$-интегралы
  %
\begin{equation}
K^q_{p,0}=(-1)^{p+q}\sum^{\min(p,q)}_{n=1} \sum^{p+q-n}_{m=n}
A^{*(1)}_{p,q;n,m}\Omega^{(n,m)}_{*ab},
\label{E2P10.7}
\end{equation}
\begin{equation}
A^{*(1)}_{p,q;n,m} \! = \! 8 \nu^{-1}_q
\mu_a^n\mu_b^m\sum_{j=j_0}^{m-n}{2^{m-j} \over
n!j!(m-j-n)!}\sum_{i=i_0}^{s-2m}
G_{pqjim},
\label{E2P10.8}
\end{equation}
%%%  Number: 1.120  %%%
%
\begin{displaymath}
G_{pqjim}= {(-1)^{i+n}\mu_b^{i}
(1-\mu_a/\mu_b)^{s-i-2m} \over (i+m-p)!(i+m-q)!(s-i-2m)!}
\times
\end{displaymath}
\begin{displaymath}
\times
\frac{\Gamma(i+m+3/2) (2(i+m)+j-s)!}{\Gamma(m+3/2)(2(i+m)-s)!},\qquad s=j+p+q.
%\label{E2P10.9}
\end{displaymath}
%Здесь
Here
\begin{displaymath}
 j_0=\max(0,m-p,m-q,2m-p-q),\qquad i_0=\max(0,p-m,q-m,(j+p+q-2m)/2).
\end{displaymath}
%
%Матричные элементы второго типа выражаются через $\Omega_*$-интегралы следующим %образом
Matrix elements of the second type are expressed in terms of $ \Omega_* $ - integrals as follows
\begin{equation}
(K^a_{a,b})^q_{0,p}=-8\nu^{-1}_q \mu^{p}_b \mu^{q}_a 
\sum^{\min(p,q)}_{n=1}\sum^{p+q-n}_{m=n}A_{p,q;n,m}^{*}
\Omega^{(n,m)}_{*ab},
\label{E0P2.27a} 
\end{equation}
%
%где
where
\begin{eqnarray}
A_{p,q;n,m}^{*(2)}=
%{(-1)^{m-n}\Gamma(p+q-n+3
%/2)(p+q-2n)!\ 2^{n}\over n!(p-n)!(q-n)!\Gamma(m+3/2)(m-n)!
%(p+q-n-m)!}=
%\nonumber\\
{2^{n}\Gamma (p+q-n+3/2)\over n!(p-n)!(q-n)!}{p+q-2n \choose m-n}
{(-1)^{m+n}\over \Gamma (m+3/2)}
\label{EOPB6} 
\end{eqnarray}
%Из этих формул понятно, что представление МЭ  второго типа  через  $\Omega_*$-%интегралы значительно проще. Поэтому естественно стартовать в рекуррентной процедуре %с МЭ второго типа.
From these formulas it is clear that the representation of ME of the second type in terms of $ \Omega_* $ - integrals is much simpler. Therefore, it is natural to start the recurrent procedure with ME of the second type.

\section{Derivation of non-linear matrix elements}

%Покажем, что стартуя с линейных МЭ второго типа (\ref{E0P2.27a}) с помощью %соотношений (\ref{E0P1.37}) можно построить все изотропные МЭ, как нелинейные, так и %линейные первого типа. 
Let us show that starting with linear ME of the second type (\ref{E0P2.27a})  and using the relations (\ref{E0P1.37}) one can construct  all isotropic MEs, both non-linear and linear of the first type.
%Несмотря на то, что в (\ref{E0P1.37}) входит производная по температуре, связи между %коэффициентами разложения МЭ по $\Omega$-интегралам, как будет показано ниже, %оказываются чисто алгебраическими. 
Despite the fact that (\ref{E0P1.37}) includes the  ME derivative with respect to temperature , the relationships between the coefficients of the ME expansion in $ \Omega $ -integrals, as will be shown below, turn out to be algebraic.

%Учитывая (\ref{E0P2.21a}), легко убедиться, что $\Omega_*$-интегралы связаны  между %собой так же как и $\Omega$-интегралы (\cite{FK})
Taking into account (\ref{E0P2.21a}), it is easy to see that $ \Omega_* $ - integrals are related to each other  in the same way as $ \Omega $ -integrals (\cite{FK})
\begin{equation}
T {\partial \Omega^{(n,m)}_{*ab}\over \partial T} + (m+3/2)
\Omega^{(n,m)}_{*ab}= \Omega^{(n,m+1)}_{*ab}.
\label{E0P2.25}
\end{equation}
%Запишем рекуррентное   соотношение (\ref{E0P1.37}) в виде
We write the recurrence relation (\ref{E0P1.37}) in the form
\begin{displaymath}
(r_1+1)(K^a_{a,b})^{r}_{r_1+1,r_2}=- (r_2+1)(K^a_{a,b})^{r}_{r_1,r_2+1}-T {d(K^a_{a,b})^{r}_{r_1,r_2}\over dT} 
\end{displaymath}
\begin{equation}
+(r_1+r_2-r)(K^a_{a,b})^{r}_{r_1,r_2}+r(K^a_{a,b})^{r-1}_{r_1,r_2}.
\label{E0P1.371}
\end{equation}
%

%В дальнейшем удобно в качестве второго нижнего индекса в матричном эдементе
% $ K^a_{a,b})^{r}_{r_1,r_2}$  использовать не $ r_2 $, а $ N=r_1+r_2$. Обозначим %соответствующий МЭ через $ K^{'r}_{r_1,N}$, т.е.
In what follows, it is convenient to use not $ r_2 $, but $ N = r_1 + r_2 $ as the second lower index in the matrix element $ (K ^ a_ {a, b}) ^ {r} _ {r_1, r_2} $. We denote the corresponding ME by $ K^{'r}_{r_1, N} $, i.e.      
\begin{equation}
K^{'r}_{r_1,N}=(K^a_{a,b})^{r}_{r_1,N-r_1}.
\label{E0P2.46} 
\end{equation}

%В этих обозначениях после замены $ N$ на   $N-1 $  (\ref{E0P1.371}) принимает вид
In this notation, after the replacement of $ N $ by $ N-1 $ (\ref{E0P1.371}) takes the form
\begin{displaymath}
(r_1+1)K^{'r}_{r_1+1,N} = -(N-r_1)K^{'r}_{r_1,N} - T {d\over dT}
K^{'r}_{r_1,N-1}+
\end{displaymath}
\begin{equation}
+(N-1-r)K^{'r}_{r_1,N-1}+rK^{'r-1}_{r_1,N-1}.
\label{E0P2.47} %! - ML - !%
\end{equation}
%%%  Number: 1.130  %%%
%При $r_1=0$ в соответствии с (\ref{E0P2.27a}) и (\ref{E0P2.46})
%запишем матрицу $K^{'}$  в виде
In accordance with (\ref{E0P2.27a}) and (\ref{E0P2.46}) for $ r_1 = 0 $ we have
\begin{equation}
K^{'r}_{0,N}=\sum^{\min (N,r)}_{n=1} \sum^{N+r-n}_{m=n} \beta^{r}_{0,N}(n,m)
\Omega ^{(n,m)}_{*ab},
\label{E0P2.48}
\end{equation}
%
%где
where
\begin{equation}
\beta^{r}_{0,N}(n,m)=8\nu^{-1}_r \mu^{N}_b \mu^{r}_a {A}_{N,r;n,m}^{*}.
\label{E0P2.49} %! - ML - !%
\end{equation}
%%%  Number: 1.132  %%%
%
%Из (\ref{E0P2.48}) видно, что $K^{'r}_{0,N}$ представляет собой линейную
%комбинацию $\Omega_*$-интегралов, где $n$ и $m$ изменяются в областях
 Thus $ K^{'r}_{0, N} $ is a linear combination of $ \Omega_* $ - integrals, where  summation indices  $ n $ and $ m $ vary in  the range
\begin{equation}
1\leq n\leq \min (N,r),\ \ \ \ \ \ n\leq m\leq N+r-n.
\label{E0P2.50} %! - ML - !%
\end{equation}
%%%  Number: 1.133  %%%

%   Покажем, что области изменения $n$ и $m$ всегда будут определяться
%формулами (\ref{E0P2.50}) и не будут зависеть от $r_1$.

%Рассмотрим рекуррентное соотношение (\ref{E0P2.47}) при $r_1=0$. Сначала, используя %(\ref{E0P2.25}), перепишем член, связанный с производной:
Let us consider the recurrence relation (\ref{E0P2.47}) for $ r_1 = 0 $. First, using (\ref{E0P2.25}), we rewrite the term containing the derivative:
\begin{displaymath}
T {d\over dT} K^{'r}_{0,N-1}= 
\sum^{\min (N-1,r)}_{n=1} \sum^{N+r-n}_{m=n+1} \beta^{r}_{0,N-1}(n,m)
\Omega ^{(n,m+1)}_{*ab} -
\end{displaymath}
\begin{equation}
-\sum^{\min (N-1,r)}_{n=1} \sum^{N-1+r-n}_{m=n} (m+3/2)
\beta^{r}_{0,N}(n,m)\Omega^{(n,m)}_{*ab}.
\label{E0P2.51} %! - ML - !%
\end{equation}
%
%Переходя в первой из двойных сумм в (\ref{E0P2.51}) от переменной суммирования $m$ к %$m-1$ и подставляя результат в (\ref{E0P2.47}), получаем
Passing in the first of the double sums in (\ref{E0P2.51}) from the summation variable $ m $ to $ m-1 $ and substituting the result in (\ref{E0P2.47}), we get
\begin{displaymath}
K^{'r}_{1,N}=
-NK^{'r}_{0,N}+(N-1-r)K^{'r}_{0,N-1}+rK^{'r-1}_{0,N-1} -
\end{displaymath}
\begin{displaymath}
- \sum^{\min(N-1,r)}_{n=1}\sum^{N+r-n}_{m=n+1}\beta^{r}_{0,N-1}(n,m-1)\Omega
^{(n,m)}_{*ab} +
\end{displaymath}
\begin{equation}
+ \sum^{\min (N-1,r)}_{n=1} \sum^{N-1+r-n}_{m=n}
(m+3/2)\beta^{r}_{0,N-1}(n,m)\Omega ^{(n,m)}_{*ab}.
\label{E0P2.52} %! - ML - !%
\end{equation}
%
%Отсюда видно, что пределы суммирования по $n$, $m$ не расширяются при %дифференцировании  $K^{'r}_{0,N}$ по температуре. Подставляя в (\ref{E0P2.52}) МЭ в %виде (\ref{E0P2.48}), получим разложение $K^{'r}_{1,N}$ по $\Omega_*$-интегралам,  
That is, the limits of summation over $ n $ and $ m $ do not extend when $ K^{'r}_{0, N} $ is differentiated  with respect to temperature. Substituting the  expression for $ K^{'r}_{0, N} $ (\ref{E0P2.48}) in (\ref{E0P2.52}), we obtain the expansion of $ K^{'r}_{1, N} $ in the $ \Omega_* $ integrals
%
%\begin{displaymath}
%K^{'r}_{1,N}= \sum^{\min (N,r)}_{n=1} \sum^{N+r-n}_{m=n}
%\beta^{r}_{1,N}(n,m) \Omega ^{(n,m)}_{*ab},
%\label{E0P2.53} %! - ML - !%
%\end{displaymath}
%
%которое имеет такую же структуру, как (\ref{E0P2.48}). 
that has the same structure as (\ref{E0P2.48}).
%Отметим, что область суммирования (\ref{E0P2.50}) при этом не изменяется. %Последовательно увеличивая $r_1$, можно таким же образом показать, что и в общем %случае пределы суммирования в ${K^{'}}^{r}_{r_1+1,N}$ не зависят от $r_1$, и %произвольный матричный элемент представляется в виде линейной комбинации 
Note that the summation  limits (\ref{E0P2.50}) does not change in this case. Consistently increasing $ r_1 $, one can show in the same way that in the general case the limits of summation in  expression for $ {K^{'}}^{r}_{r_1 + 1, N} $ do not depend on $ r_1 $, and  arbitrary matrix element is represented as a linear combination
\begin{equation}
K^{'r}_{r_1,N}= \sum^{\min (N,r)}_{n=1} \sum^{N+r-n}_{m=n}
\beta^{r}_{r_1,N}(n,m)\Omega ^{(n,m)}_{*ab},
\label{E0P2.55} 
\end{equation}
%
%Понятно, что такая же формула с другими коэффицинтами справедлива и при разложении по $\Omega$-интегралам. 
It is clear that the same formula with other coefficients is also valid for the expansion in $ \Omega $ -integrals.

%Для $n>\min(N,r)$ и $m>N+r-n$ доопределим матрицу $\beta^r_{r_1,N}(n,m)$ нулями.
 Let us extend the matrix $ \beta^r_ {r_1, N} (n, m) $
 setting the elements with $ n> \min(N, r) $ and $ m> N + r-n $ equal to zero. 
%
%Тогда можно записать
Then we can write
\begin{displaymath}
K^{'r}_{r_1,N}= \sum^\infty_{n=1} \sum^\infty_{m=n}
\beta^{r}_{r_1,N}(n,m) \Omega ^{(n,m)}_{*ab}.
\end{displaymath}
%
%Подставляя последнее равенство в (\ref{E0P2.47}) и приравнивая коэффициенты при одинаковых $\Omega_*$-интегралах, получим следующее рекуррентное соотношение, где с помощью $\Theta$-функций явно выделены области ненулевых значений $\beta^r_{r_1,N}(n,m)$
Substituting the last equality in (\ref{E0P2.47}) and equating the coefficients for the same $ \Omega_* $ integrals, we obtain the following recurrence relation, where $ \theta $ -functions explicitly identify ranges of indices where $ \beta^r_ {R_1, N} (n, m) $ are non-zero
\begin{displaymath}
\beta^{r}_{r_1+1,N}(n,m) ={1\over r_1+1}\bigl[-(N-r_1)
\times
\end{displaymath}
\begin{displaymath}
\times
\Theta (\min (N,r)-n)\Theta (N+r-n-m)
\beta^{r}_{r_1,N}(n,m)+(m+N-1-r+3/2)
\times
\end{displaymath}
\begin{displaymath}
\times
\Theta (\min (N-1,r)-n)\Theta(N-1+r-n-m)
\beta^{r}_{r_1,N-1}(n,m)+
\end{displaymath}
\begin{displaymath}
+r\Theta (\min (N-1,r-1)-n)\Theta (N+r-2-n-m)\beta^{r-1}_{r_1,N-1}(n,m)-
\end{displaymath}
\begin{equation}
-\Theta(\min(N-1,r)-n)\Theta(N+r-n-m)(1-\delta_{m,n})
\beta^r_{r_1,N-1}(n,m-1) \bigr].
\label{E0P2.60} %! - ML - !%
\end{equation}
%%%  Number: 1.138  %%%
%
%Эта же рекуррентная формула справедлива и при разложении по $\Omega$-интегралам. Различие будет только в начальных коэффициентах разложения $\beta^{r}_{0,N}(n,m)$. 
The same recurrence formula is  valid for the expansion in $ \Omega $ -integrals.  The only difference is in the initial expansion coefficients$ \beta^{r}_{0, N} (n, m) $.
%Особо отметим универсальность рекуррентной формулы (\ref{E0P2.60}) для матрицы $\beta$. При любом фиксированном отношении масс она не зависит от сечения взаимодействия.
We especially note  that for any fixed mass ratio the recurrence formula (\ref{E0P2.60}) for the matrix $ \beta $  is universal and does not depend on the interaction cross section.

%Нами была составлена программа расчета матрицы $\beta$ по формуле (\ref{E0P2.60}). Получено полное совпадение коэффициентов разложения линейных изотропных МЭ первого типа $(K^a_{a,b})^q_{0,p}$ по $\Omega$-интегралам, с аналитически найденными в \cite{Mas1} при нескольких первых значениях индексов. Подчеркнем, что  в нашем подходе такие МЭ вычисляются в самом конце рекуррентной процедуры после определения всех нелинейных МЭ.
We wrote a program for calculating the matrix $ \beta $ by the formula (\ref {E0P2.60}). The  obtained coefficients of the expansion of the linear isotropic ME of the first type $ (K^a_{a, b})^q_{0, p} $  in $ \Omega $ -integrals,  coincide with ones that analytically found in \cite{Mas1}  for several first indices. We emphasize that  these  MEs are calculated at the  final step of the recurrence procedure after the determination of all non-linear ME.

%Таким образом, показано, что  связи между МЭ могут быть использованы как рекуррентные соотношения при вычислении нелинейных МЭ для произвольных законов взаимодействия.
Thus, it is shown that the connections between MEs can be used as recurrence relations in the calculation of non-linear ME for arbitrary interaction laws.

%Отметим, что расчет коэффициентов разложения $ \beta^r_{r_1,N}(n,m) $ должен проводиться с высокой точностью. Для реализации расчета с учетом большого количества значащих цифр мы использовали пакет MPFUN2015 \cite{precise}. Было проведено сравнение изотропных нелинейных МЭ $K^r_{r_1,r_2}$ (15 значащих цифр) с индексами в диапазоне $[0, 32]$, рассчитанных с помощью формул (\ref{E0P2.55}), (\ref{E0P2.60}) и найденных с помощью рекуррентной формулы (\ref{recrelp}) для случая твердых шаров. Получено полное совпадение результатов. 
Note that the calculation of the expansion coefficients $ \beta^r_ {r_1, N} (n, m) $ should be carried out with high accuracy. To implement the calculation, taking into account a large number of significant digits, we used the package MPFUN2015 \cite{precise}.  We compared isotropic non-linear ME $ K^r_{r_1, r_2} $ (15 significant digits) with indices in the range $ [0, 32] $ calculated using the formulas (\ref{E0P2.55}), (\ref{E0P2.60}) and  determined by the recurrence  relations (\ref{recrelp}) for the case of hard spheres. A complete coincidence of the results  has been observed.

\section{Power interaction law }

%Найдем выражение для $\Omega ^{(n,m)}_{*ab}$ в случае степенных законов взаимодействия. Подставляя сечение (\ref{ppsigma}) в выражение для $Q_{*ab}^{(n)}$ (\ref{E0P2.16}), после  интегрирования по $\bar{g}$ получаем 
Let us  derive the expression for $ \Omega^{(n, m)}_{* ab} $ in the case of power interaction law. Substituting the cross section (\ref{ppsigma}) into the expression for $ Q_{* ab}^{(n)} $ (\ref{E0P2.16}), after integrating over $ \bar {g} $, we get
\begin{equation}
\Omega ^{(n,m)}_{*ab}= 2^{\mu-2}\pi^{-1/2}\left({kT\over m_{ab}}\right)^
{\mu}\ 2^{n}J_{n}(\mu) \Gamma (m+3/2+\mu),
\label{E0P2.33}
\end{equation}
%
%где величина ${J}_{n}(\mu)$, в отличие от случая произвольных потенциалов, связана только с угловой частью сечения: 
 Here $ {J}_{n} (\mu) $, unlike the case of arbitrary potentials, is  determined by the angular part of the cross section only:
\begin{equation}
{J}_{n}(\mu) = 4\pi \int^1_{0} F^{ab}_\gamma(z) z^{n}\ dz .
\label{E0P1.63}
\end{equation}
%%%  Number: 1.61  %%%
%

%Покажем, что в этом случае выражение для линейных изотропных МЭ второго рода, которые являются стартовыми в рекуррентной процедуре, основанной на (\ref{recrelp}), оказываются существенно проще, чем в общем случае. Для этого подставим представление $\Omega_{*ab} $ (\ref{E0P2.33})  в разложение (\ref{E0P2.27a})--(\ref{EOPB6}):
Let us show that in this case the expression for linear isotropic ME of the second kind, which are the starting ones in the recurrence procedure based on (\ref{recrelp}), are much simpler than in the general case. To do this, we substitute the 
$ \Omega_{* ab} $ (\ref{E0P2.33}) into the expansion (\ref{E0P2.27a}) - (\ref{EOPB6}):
 \begin{displaymath}
K_{0,p}^q=-{2^{\mu+1}\over \pi^{1/2} \nu_q}
\left({kT\over m_{ab}}\right)^{\mu}
\!\!\mu^{p}_b \mu^{q}_a\!\!\sum^{\min(p,q)}_{n=1}\!\! J_{n}(\mu)
{2^{2n}\Gamma (p+q-n+3/2)\over n! (p-n)! (q-n)!}
\times
\end{displaymath}
\begin{equation}
\times
\sum^{p+q-2n}_{s=0}{p+q-2n\choose s}(-1)^s{\Gamma(n+s+3/2+\mu)
\over \Gamma (n+s+3/2)}.
\label{E0P2.38} 
\end{equation}
%Используя  интегральное представление $ B $-функции \cite{GR}, можно показать, что 
Using the integral representation of the $ B $ -function \cite{GR}, it can be shown that
\begin{equation}
\sum\limits_{k=0}^{t}
{t\choose k}(-1)^k\frac{\Gamma(a+k)}{\Gamma(b+k)}=
\frac{\Gamma(a)\Gamma(t+b-a)}{\Gamma(b-a)\Gamma(t-b)}\quad a>0, \\ b>0.
\label{beta}
\end{equation}
% Заменяя  последнюю сумму в (\ref{E0P2.38})  выражением  (\ref{beta}) при $ t=p+q-2n $, $ a= n+3/2+\mu$, и $ b=n+3/2$, получаем  представление линейных изотропных МЭ второго типа в виде однократной суммы 
 Replacing the last sum in (\ref{E0P2.38}) by the expression (\ref{beta}) for $ t = p + q-2n $, $ a = n + 3/2 + \mu $, and $ b = n + 3/2 $, we obtain  the linear isotropic ME of the second type in the form of a single sum
\begin{equation}
K^q_{0,p}=\left({2kT\over m_{ab}}\right)^\mu
\mu^p_b \mu^{q}_a {q!\over \Gamma(q+3/2)}
\sum^{\min(p,q)}_{n=1}J_{n}{2^{2n}\Gamma(n+3/2+
\mu)\Gamma (p+q-2n-\mu)\over n!(p-n)!(q-n)!\Gamma (-\mu)}.
\label{eop2.40} %
\end{equation}
%Таким образом, в случае степенных законов взаимодействия получена простая формула для линейных изотропных МЭ второго типа с произвольным соотношением масс. Эта формула для простого газа ($\mu_a=\mu_b=1/2$, $m_{ab}=m/2$) совпадает с выражением, полученным ранее другим способом в \cite{PhFl, book}. 
Thus, in the case of power interaction law, a simple formula is obtained for linear isotropic ME of the second type with an arbitrary mass ratio. This formula for a monoatomic gas ($ \mu_a = \mu_b = 1/2 $, $ m_ {ab} = m / 2 $)  coincide with the expression obtained earlier  by other method in \cite{PhFl, book}.

%Выразим $J_{n}$ через параметры степенного потенциала. Из  (\ref{E0P2.10}) следует, что для степенных потенциалов 
 Let us express $ J_{n} $ in terms of the parameters of the power potential. It follows from (\ref{E0P2.10}) that for power potentials
\begin{equation}
Q^{(n)}_{*ab}=
%\left({2kT\over m_{ab}}\right)^{\mu-1/2}\bar
{g}^{2\mu -1}
\ 2^n J_n(\mu).
\label{E0P2.30}
\end{equation}
%
%Рассмотрим потенциал   
Consider the potential
\begin{equation}
\varphi=\left( \frac{\epsilon}{r} \right)^\nu,\qquad \nu=\frac{4}{1-2\mu} .
\label{phi}
\end{equation}
%
%Известно (см., например, \cite{FK}), что в случае потенциала (\ref{phi}) $ Q^{(n)}_{ab} $ (\ref{E0P2.10}) можно представить в виде 
It is known (see, for example, \cite{FK}), that in the case of the potential (\ref{phi}) $ Q^{(n)}_{ab} $ (\ref{E0P2.10}) can be represented in the form
\begin{equation}
Q^{(n)}_{ab}=2\pi\epsilon^2(\frac{m_{ab}g^2 }{2\nu})^{-2/\nu}A_n(\nu),
\label{qab}
\end{equation}
%где
where
\begin{equation}
A_n(\nu)=\int^\infty_{0}(1-\cos^n(\theta(s,\nu)))s ds.
\label{annu}
\end{equation}
%Способ построения угла $\theta(s,\nu)$ можно найти, например, в \cite{FK}. Там же приведены значения $ A_n(\nu) $ при  $ n=1,2 $ для ряда целых значений $\nu $ из области    $ 4\leq\nu\leq \infty$.
The method of constructing the angle $ \theta (s, \nu) $ can be found, for example, in \cite{FK}. The values of $ A_n (\nu) $ for $ n = 1,2 $ for a  set of integer values $ \nu $ in the range $ 4 \leq \nu \leq \infty $ are also given there.

%Аналогичная формула связывает $ Q^{(n)}_{* ab} $ и величину
An analogous formula  is valid for $ Q^{(n)}_{* ab} $ 
\begin{equation}
Q^{(n)}_{* ab}=2\pi\epsilon^2(\frac{m_{ab}g^2 }{2\nu})^{-2/\nu} \tilde A_n(\nu),
\label{qab1}
\end{equation}
%где
where
\begin{equation}
\tilde A_n(\nu)=\int^\infty_{0}(1-\cos(\theta(s,\nu)))^nsds.
\label{annu}
\end{equation}
%
%Из (\ref{E0P2.30}), представляя $ Q^{(n)}_{*ab} $ в виде, аналогичном (\ref{qab}), получаем 
Substituting (\ref{qab1}) to (\ref{E0P2.30}) we get
\begin{equation}
J_n(\mu)=
  \pi\epsilon^2\left( \frac{m_{ab} }{2\nu} \right)^{-2/\nu}\tilde A_n(\nu) .
\label{cfi1}
\end{equation}
%
%Таким величина $ J_n(\mu) $ под знаком суммы в (\ref{eop2.40}) полностью определена. 
Thus, the value $ J_n(\mu) $  in (\ref{eop2.40}) is completely defined.

\section{Conclusion} 

%При расчете сильно неравновесных ФР моментным методом необходимо учитывать большое число членов в разложении ФР, следовательно, необходимы матричные элементы с большими индексами. В наших работах \cite{PhFl,book} показано, что МЭ с соседними индексами связаны соотношениями, которые следуют из инвариантности интеграла столкновений относительно выбора параметров весового максвеллиана в разложении по функциям Барнетта, а именно, температуры и средней скорости. 
When calculating highly non-equilibrium distribution functions by the moment method, it is necessary to take into account a large number of terms in the expansion of the distribution function, hence, matrix elements with large indices are necessary. In our papers \cite{PhFl, book} it is shown that MEs with neighbouring indices are connected by the relations that follow from the invariance of the collision integral with respect to the choice of the Maxwellian weight parameters in the Barnett expansion, namely, temperature and mean velocity.

%В случае степенных законов взаимодействия все соотношения, связывающие МЭ, являются алгебраическими. Это позволило построить рекуррентную процедуру для последовательного определения МЭ \cite{PhFl,book}. Найденные МЭ были использованы для расчета  ФР при сильном отклонении от равновесия, в частности, в сильных постоянных и переменных внешних полях \cite{ender2009}. 
In the case of power interaction law, all the relations connecting ME are algebraic. This allowed us to construct a recurrence procedure for sequential determination of the ME \cite{PhFl, book}. The  obtained matrix elements were used to calculate the distribution function  in the case of strong deviation from equilibrium, in particular, in strong constant and variable external fields \cite{ender2009}.

%Для произвольных законов взаимодействия соотношения связывающие неизотропные МЭ (они следуют из инвариантности относительно выбора средней скорости весового максвеллиана), также являются алгебраическими. На их основе в \cite{RecME1} была построена рекуррентная процедура, которая позволяет найти все неизотропные МЭ, если известны изотропные МЭ. Таким образом, построение изотропных МЭ приобретает особое значение. 
For arbitrary laws of interaction, the relations connecting non-isotropic MEs (they follow from invariance with respect to the choice of the mean velocity of the weight Maxwellian) are also algebraic. On their basis, a  reccurence procedure was constructed in \cite{RecME1}, which allows finding all non-isotropic MEs if isotropic MEs are known. Thus, the construction of isotropic ME acquires special significance.

%Для произвольных (не степенных) законов взаимодействия соотношения, связывающие изотропные МЭ, включают производную по температуре. Переходу к алгебраическим соотношениям помогает использование хорошо известных в кинетической теории $ \Omega $-интегралов. Учитывая, что производная от $ \Omega $-интеграла по температуре выражается через линейную комбинацию 
%$ \Omega $-интегралов, удается записать алгебраические соотношения для коэффициентов разложения МЭ по $ \Omega $-интегралам.
For arbitrary (not power-law) laws of interaction, the relations connecting isotropic ME include a derivative with respect to temperature. The use of $ \Omega $ -integrals well-known in the kinetic theory  allows to reduce the relations to algebraic  ones. Taking into account that the derivative of the $ \Omega $ -integral with respect to temperature is expressed in terms of a linear combination of $ \Omega $ -integrals, it is possible to obtain algebraic relations for the coefficients of the ME expansion in $ \Omega $ -integrals. 

%При этом оказалось, что более удобно рассматривать коэффициенты разложения по $ \Omega_*$-интегралам (\ref{E0P2.17}), которые легко выражаются через $\Omega$-интегралы. Эти коэффициенты последовательно определяются с помощью описанной в настоящей статье рекуррентной процедуры, причем в качестве стартовых целесообразно использовать коэффициенты разложения линейных изотропных МЭ второго типа $ K^r_{0,r_2} $, которые имеют наиболее простой вид (\ref{EOPB6}). Необходимо отметить, что соотношения, связывающие коэффициенты разложения, являются общими для $ \Omega_*$ и $ \Omega$-интегралов, они также не зависят от масс сталкивающихся частиц. Отношение масс входит только в стартовые коэффициенты разложения (\ref{EOPB6}).  
It turned out that it is more convenient to consider the coefficients of the expansion in $ \Omega_* $ - integrals (\ref{E0P2.17}), which are easily expressed in terms of $ \Omega $ -integrals. These coefficients are sequentially determined using the recurrence procedure described in this paper, and it is advisable to use the expansion coefficients of linear isotropic ME of the second type $ K^r_{0, r_2} $, which have the simplest form (\ref{EOPB6}). It should be noted that the relations connecting the expansion coefficients are common for $ \Omega_* $ and $ \Omega $ -integrals, they also do not depend on the masses of the colliding particles. The mass ratio enters only in the initial expansion coefficients (\ref{EOPB6}).

%Нами была реализована процедура, позволяющая последовательно вычислять все изотропные МЭ с использованием в качестве входных данных набора известных (табулированных) $\Omega$-интегралов. Она требует вычислений с повышенной точностью, для чего был использован  пакет MPFUN2015 \cite{Precise}. Программа была протестирована на примере вычисления МЭ для модели твердых шаров.
We have implemented a procedure that allows us to compute sequentially all isotropic MEs using  a set of known (tabulated) $ \Omega $ -integrals. It requires computations with increased accuracy, for which the MPFUN2015 \cite{precise} package was used. The program was tested by the ME calculation for the hard sphere model. 

%Для степенных законов взаимодействия представление линейных изотропных МЭ второго типа в виде разложения по $\Omega_*$-интегралам  позволило обобщить полученную в \cite{PhFl,book} формулу на случай произвольного соотношения масс. Их разложение при этом сводится к однократной сумме. 

 For power interaction laws, the expansion of linear isotropic ME of the second type in $ \Omega_* $ integrals
is reduced to a single sum. This representation made it possible to generalize the formula obtained in \cite{PhFl, book}  to the case of an arbitrary mass ratio. 

%Для произвольных законов взаимодействия изотропные нелинейные МЭ могут быть рассчитаны по формуле (\ref{E0P2.55}), в которой  коэффициенты $\beta^{r}_{r_1+1,N}(n,m) $ определяются с помощью рекуррентной процедуры. В свою очередь, как было указано выше, они являются стартовыми для расчета неизотропных нелинейных МЭ.
For arbitrary  interaction laws, isotropic non-linear ME can be calculated by the formula (\ref {E0P2.55}), in which the coefficients $ \beta^{r}_{r_1 + 1, N} (n, m) $ are determined by the use of a  recurrence procedure. In turn, as mentioned above, they are the starting  values for calculating non-isotropic non-linear ME.

%Таким образом, совместное применение описанных в этой статье и в нашей предыдущей статье \cite{RecME1} рекуррентных процедур дает возможность найти все МЭ нелинейного интеграла столкновений для произвольных законов взаимодействия и произвольных масс сталкивающихся частиц. Табулированные МЭ могут затем использоваться для решения различных задач. Это существенно расширяет возможности моментного метода, который может теперь применяться для расчета сильно неравновесных ФР в реальных физических ситуациях. 
Thus, the  use of the both recurrence procedures described in this article and in our previous article \cite{RecME1} makes it possible to find all the ME of the non-linear collision integral for arbitrary laws of interaction and arbitrary masses of the colliding particles. Tabulated matrix elements can then be used to solve various  problems. This greatly expands the application of the moment method, which can now be used to calculate highly non-equilibrium distribution functions in real physical situations.

\end{spacing}

\end{document}